\documentclass[conference]{IEEEtran}
\IEEEoverridecommandlockouts
\usepackage[utf8]{inputenc}

\usepackage{tikz} 
\usetikzlibrary{backgrounds}
\usepackage{pgfplots} 
\usepackage{pgfplots}
\usepackage{siunitx}
\usepackage{cite}
\usepackage{pbox}
\usepackage{rotating}
\usepackage{tabularx}
\usepackage{afterpage}
\usepackage{amsmath,amssymb,amsfonts}

\usepackage{graphicx}
\usepackage{float}
\usepackage{mathtools, cuted}
\usepackage{textcomp}
\usepackage{verbatim}
\usepackage[overload]{textcase}
\usepackage{verbatim}
\usepackage{gensymb}
\usepackage{algorithm}
\usepackage{algpseudocode}
\DeclareMathOperator*{\argmax}{arg\,max}
\DeclareMathOperator*{\argmin}{arg\,min}
\makeatletter
\newcommand{\algmargin}{\the\ALG@thistlm}
\makeatother
\newlength{\whilewidth}
\algdef{SE}[parWHILE]{parWhile}{EndparWhile}[1]
  {\parbox[t]{\dimexpr\linewidth-\algmargin}{%
     \hangindent\whilewidth\strut\algorithmicwhile\ #1\ \algorithmicdo\strut}}{\algorithmicend\ \algorithmicwhile}%
\algnewcommand{\parState}[1]{\State%
  \parbox[t]{\dimexpr\linewidth-\algmargin}{\strut #1\strut}}

\title{Scheduling for Ground-Assisted Federated Learning in LEO Satellite Constellations}
\author{\IEEEauthorblockN{Nasrin Razmi\IEEEauthorrefmark{1}\IEEEauthorrefmark{2}, Bho Matthiesen\IEEEauthorrefmark{1}\IEEEauthorrefmark{2}, Armin Dekorsy\IEEEauthorrefmark{1}\IEEEauthorrefmark{2}, and Petar Popovski\IEEEauthorrefmark{3}\IEEEauthorrefmark{1}}

\IEEEauthorblockA{\IEEEauthorrefmark{1}University of Bremen, U Bremen Excellence Chair, Dept.\ of Communications Engineering, Germany}

\IEEEauthorblockA{\IEEEauthorrefmark{2}Gauss-Olbers Center c/o University of Bremen, Dept. of Communications Engineering, Germany}

\IEEEauthorblockA{\IEEEauthorrefmark{3}Department of Electronic Systems, Aalborg University, Aalborg, Denmark}

\text{Email:\{razmi, matthiesen, dekorsy\}@ant.uni-bremen.de, petarp@es.aau.dk}
\thanks{This work was funded in part by
by the German Research Foundation (DFG) under Germany's Excellence Strategy (EXC 2077 at University of Bremen, University Allowance).}

}

\newcommand{\nas}[1]{{\textcolor{black}{#1}}}

\begin{document}

\maketitle
\begin{abstract}

Distributed training of machine learning models directly on satellites in low Earth orbit (LEO) is considered. Based on a federated learning (FL) algorithm specifically targeted at the unique challenges of the satellite scenario, we design a scheduler that exploits the predictability of \nas{visiting} times between ground stations (GS) and satellites to reduce model staleness. Numerical experiments show that this can improve the convergence speed by a factor three.

\end{abstract}

\vspace{-0.31 cm}
\section{Introduction}
The small low Earth orbit~(LEO) satellites, efficient in terms of cost and deployment, are marking a new era in satellite communications, as well as their integration with terrestrial \nas{networks}. These LEO satellites are commonly deployed in large constellations, thereby creating a moving infrastructure for a seamless global coverage for, various applications such as communication services and Earth observation~\cite{5G-access}. Many of these applications are data-intensive. For instance, in Earth observation, the high spatial, spectral, and temporal resolution of the imaging equipment, leads to large amounts of collected data~\cite{9378798}. These data are used in a variety of applications, such as 
disaster prevention, environmental monitoring, and urban planning. Transmitting such a large amount of data to the Earth may not be practical due to scarcity of the radio frequency resources or stringent delay requirements~\cite{curzi2020large}.

\begin{figure}[t]
    \centering
    \includegraphics[height=5cm, width=8cm]{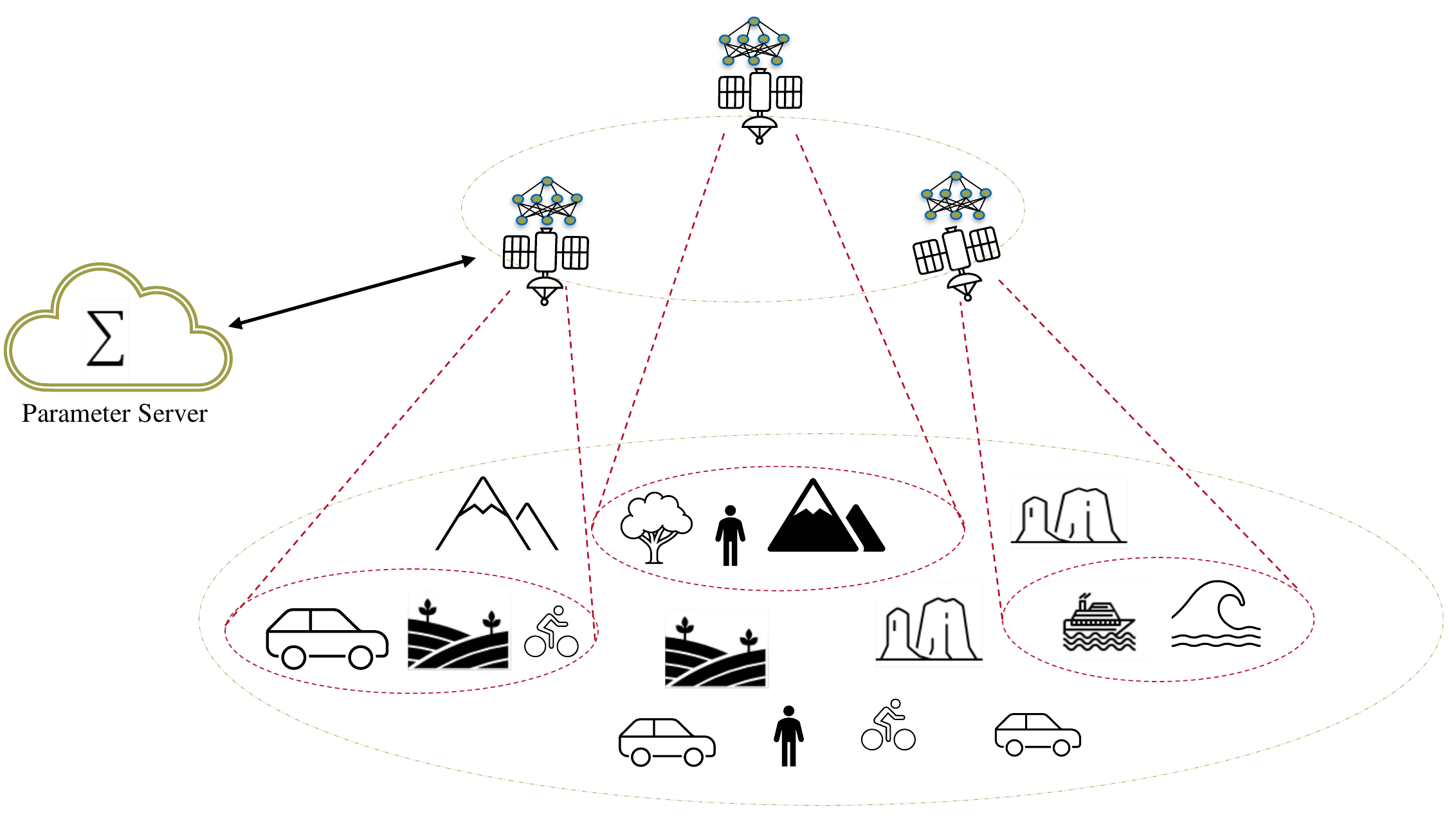}
    \caption{FL for Earth observation applications.}
    \vspace{-0.30 cm}
    \label{fig: Ground-assisted FL}
\end{figure}

To address these constraints, a plausible solution is to process the data directly on-board of the satellites and only transmit the abstracted information to the ground station~(GS). In this regard, it is viable to use federated learning~(FL) \cite{mcmahan2017communication}, as a cooperative machine learning~(ML) scheme in which the  satellites only need to transmit the model parameters to the server instead of the raw data. Fig.~\ref{fig: Ground-assisted FL} depicts the usage of FL for Earth observation. In the original FL setup, the user's participation in the training process is intermittent and randomized, based on user activity and communication availability. In satellite scenarios, the link (un)availability is predictable and related to the visiting pattern of the satellites to the location at which the GS is positioned. This GS orchestrated FL scenario was first identified in \cite{9674028}, where a new asynchronous FL procedure is proposed that addresses the unique challenges of FL on satellites. In \cite{razmi2021board}, it is shown that the presence of inter-satellite links (ISL) improves the convergence speed of FL on satellites considerably. Without this capability, several extensions of the approach in \cite{9674028} are possible that might lead to faster convergence speed. Indeed, the authors of \cite{so2022fedspace} consider a heuristic GS update procedure and gradient buffering to speed up convergence. 

In this paper, we take a different approach and design a scheduler based on the predictability of GS-satellite \nas{visiting} lengths to reduce the model staleness. This leads to significantly faster convergence speed than the baseline approach in \cite{9674028} and can be combined with different aggregation rules, such as the one in \cite{so2022fedspace}.

\vspace{-0.4 cm}
\section{System Model}

We consider a LEO satellite constellation with $P$ circular orbits where the $p$-{th} orbit contains $K_p$ satellites. The set $\mathcal K = \{S_1, S_2, \dots, S_K\}$ denotes all $K=\sum_{p=1}^{P}{K_p}$ satellites.
The altitude and inclination of orbit $p$ are denoted by $h_p$ and $i_p$, respectively. Let $T_p = \frac{2\pi(r_E + h_p)}{v_p}$ and $v_p = \sqrt{\frac{\mu}{h_p+r_E}}\,\si{\meter/\second}$, respectively, denote the orbital period and the speed of the satellites in orbit $p$ where $r_E = 6371\,\si{\km}$ is the Earth radius and $\mu = 3.98 \times 10^{14}\,\si{\meter^3/\second^2}$ is the geocentric gravitational constant.
A satellite can communicate with the GS if it is in a \emph{visiting state}, i.e., the line-of-sight link between it and the GS is not blocked by the Earth. Otherwise, it is in a \emph{ non-visiting state}. The line-of-sight link between the satellite $k$ and the GS is available when $\frac{\pi}{2} - \angle (\vec r_{GS}, \vec r_k - \vec r_{GS}) \ge \alpha_e$, where $\vec r_{k}$ and $\vec r_{GS}$ denote the position of satellite $k$ and the GS, respectively, and $\alpha_e$ is the minimum elevation angle.

We define two time instants of significance. The \emph{rise time} of the $k$-th satellite $t_{r,k}^{n}$ is the time instant at which the satellite enters in its $n$-th visit, while the \emph{set-time} $t_{s,k}^{n}$ is the time instant at which the satellite finishes its $n$-th visit. 
\nas{The rise-time sequences of all $K$ satellites, $\tau_{\mathrm{rise}}$, is expressed as}
\vspace{-0.2 cm}
\begin{equation}
    \tau_{\mathrm{rise}} = \left( \left\{t_{r,1}^{n}\right\}_{n=1}^{N_1}, \left\{t_{r,2}^{n}\right\}_{n=2}^{N_2}, ... ,\left\{t_{r,K}^{n}\right\}_{n=K}^{N_K}\right),
\end{equation}
where $t_{r,k} = \left\{t_{r,k}^{n}\right\}_{n=1}^{N_k}$ is the rise-time sequence of the $k$-th satellite with $N_k$ being the number of visiting states of the $k$-{th} satellite in the considered time interval $[T_b, T_f]$. Without loss of generality, we assume both $T_b$ and $T_f$ are located in the off-time of all satellites.
\nas{Similarly, the set-time sequences of all $K$ satellites, $\tau_{\mathrm{set}}$, is given by}
\begin{equation}
    \tau_{\mathrm{set}} = \left( \left\{t_{s,1}^{n}\right\}_{n=1}^{N_1}, \left\{t_{s,2}^{n}\right\}_{n=2}^{N_2}, \dots, \left\{t_{s,K}^{n}\right\}_{n=K}^{N_K}\right),
\end{equation}
where $t_{s,k} = \left\{t_{s,k}^{n}\right\}_{n=1}^{N_k}$ is the set-time sequence of the $k$-th satellite.

Next, we define two types of time intervals associated with these sequences. The \emph{on-time} is the time interval between the rise- and set-time, which corresponds to the duration of a visiting state. The \emph{off-time} is a time interval between two visiting states. \nas{On-time sequence for all $K$ satellites, $\tau_{\mathrm{on}}$, is defined as} 
\vspace{-0.2 cm}
\begin{equation}
\tau_{\mathrm{on}} = \left(\left\{[t_{r,1}^{n},t_{s,1}^{n}] \right\}_{n=1}^{N_1} , ..., \left\{[t_{r,K}^{n},t_{s,K}^{n}]\right\}_{n=1}^{N_K} \right),
\end{equation}
where $t_{\mathrm{on},k} = \left\{[t_{r,k}^{n},t_{s,k}^{n}]\right\}_{n=1}^{N_1}$ denotes the on-time sequence of the $k$-th satellite. \nas{Off-time sequences of the satellites, $\tau_{\mathrm{\mathrm{off}}}$, is also expressed as}
\begin{equation}
\tau_{\mathrm{\mathrm{off}}} = \left( \left\{t_{\mathrm{off},1}^{n}\right\}_{n=1}^{N_1+1}, \dots, \left\{t_{\mathrm{off},k}^{n}\right\}_{n=1}^{N_K+1}  \right)
\end{equation}
where
\begin{equation}
 t_{\mathrm{off},k}=\left\{t_{\mathrm{off},k}^{n}\right\}_{n=1}^{N_k+1} = \left\{[T_b, t_{r,k}^{1}],[t_{s,k}^{1},t_{r,k}^{2}], 
\dots, [t_{s,k}^{N_k}, T_f]   \right\}.
\end{equation}

The state $E(t,k)$ denotes the visiting state of the $k$-{th} satellite at time instant $t$, defined as 
\begin{equation} \label{eq: status}
E(t,k) =
\begin{cases}
 E_\mathrm{on}, & t \in t_\mathrm{on,k} \\
 E_\mathrm{off} , & t \in t_\mathrm{off,k} 
\end{cases}
\end{equation}
where $E_\mathrm{on}$ and $E_\mathrm{off}$ refer \nas{to} on-time and off-time states, respectively. For simplicity, in the following we remove the time component $t$ and denote the state of satellite \nas{$k$} by $E(k)$.
\vspace{-0.3 cm} 
\subsection{Computation Model}
Each satellite $k$ gathers a local dataset $\mathcal D_k = \{\pmb{x}_1,...,\pmb{x}_{D_k}\}$ from the Earth where ${\pmb{x}}_i$ and $D_k$ denote the $i$-{th} sample and the number of samples of this satellite, respectively. This data is used to train a ML model in which each satellite $k$ builds a loss function $F_k(\pmb{w})$ expressed as
\vspace{-0.1 cm}
\begin{equation} \label{eq: loss function of satellite k}
    F_k(\pmb{w}) = \frac{1}{D_k} \sum\nolimits_{\pmb{x}\in\mathcal D_k} f_k(\pmb{x}, \pmb{w}),
\end{equation}
where $f_k(\pmb{x}, \pmb{w})$ is the per-sample loss function at satellite $k$ and builds upon the learning target which can be any convex or non-convex function. The vector $\pmb{w}$ denotes the parameter describing the model.

\begin{figure*}
    \centering
    \includegraphics[width=\textwidth,height=3.5cm]{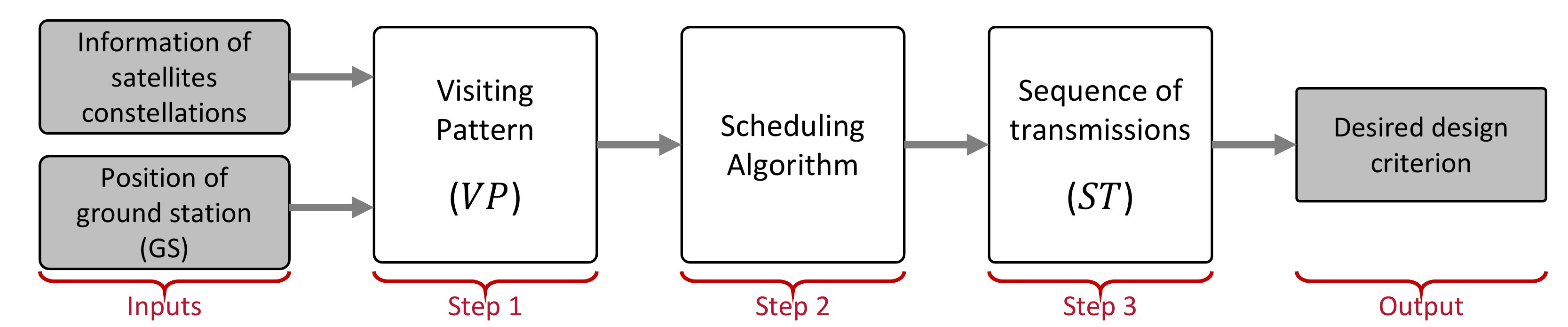}
    \caption{Satellite scheduling Algorithm.}
    \vspace{-0.2 cm}
    \label{fig: Schedule diagram for satellites}
    \vspace{-0.2 cm}
\end{figure*}

\nas{Lo}cal raw dataset of each satellite is kept private, i.e., it is \nas{shared} neither with other satellites nor with the GS. \nas{S}atellites aim to collaboratively minimize a global loss function\vspace{-0.1 cm}
\begin{equation}
    F(\pmb{w}) = \sum\nolimits_{k\in\mathcal K} \frac{D_k}{D} F_k(\pmb{w}),
\end{equation}
where $D = \sum_{k\in\mathcal K} D_k$ is the total number of \nas{s}amples.
Unlike the well-known FedAvg algorithm \cite{mcmahan2017communication}, which has only one counter for the global epoch, we \nas{define counters for the GS and each satellite}. The global epoch of the model is denoted as $n$ \nas{which} kept track by the GS. In addition, \nas{for any }satellite $k$\nas{, a local counter $n_k$ is defined to} track \nas{the} satellite \nas{participation}.
For example, in a scenario with synchronous FL and full client participation, $n_k = n$ for all $k$. We further define $\pmb{w}^{n}$ to be the global model parameters at epoch $n$.
Assuming each satellites trains the model locally for $I$ iterations using stochastic gradient descent~(SGD), the local model parameters of satellite $k$ at iteration $i\ge 1$ are
\begin{equation} \label{eq: SGD}
    {\pmb{w}_k^{n_k,i}} = {\pmb{w}_k^{n_k, i-1}} -\eta\bigtriangledown{F_k({\pmb{w}_k^{n_k, i-1}})}, 
\end{equation}
where ${\pmb{w}_k^{n_k,0}}$ is the global \nas{m}odel received by satellite $k$ in its $n_k$-{th} update and $\eta$ is the learning rate. 
Following the linear computation time model from \cite{8737464}, the time $t_l(k)$ required by satellite $k$ to compute an update to the global model is 
\begin{equation} \label{eq: Time of learning}
    t_l(k) = \frac{c_k I S(D_k)}{\nu_k},
\end{equation}
where $c_k$ is the number of CPU cycles required to process a single data bit, $S(D_k)$ is the size of data in bits, and $\nu_k$ is the CPU frequency.
\vspace{-0.2 cm}
\subsection{Communication Model}
Communication between a satellite and the GS is possible if the line of sight between them is not blocked by the Earth, i.e., satellite $k$ is in the on-time period with $E(k) = E_\mathrm{on}$. Then, the signal to noise ratio~(SNR) between the $k$-{th} satellite and the GS is \nas{written} as \cite{ippolito2017satellite}
\begin{equation} \label{eq: SNR}
\mathrm{SNR}(k,GS) =
    \begin{cases} 
     \frac{P_t G_k G_{GS} }{N_0  L(k,GS)},~&\text{if}~ E(k) = E_\mathrm{on} \\
    0,~&\text{if}~ E(k) = E_\mathrm{off},
    \end{cases}
\end{equation}
where $P_t$ is the transmission power, $G_k$ and $G_{GS}$ are the average antenna gains of satellite $k$ towards GS and vice versa, $N_0 = k_B T B$ is the total noise power with $k_B = 1.380649 \times 10^{-23}\,\si{\joule/\kelvin}$ being the Boltzmann constant, $T$ is the receiver noise temperature, and $B$ is the channel bandwidth. 
\nas{F}ree space path loss $L(k,GS)$ between the $k$-th satellite and the GS is expressed as
\vspace{-0.2 cm}
\begin{equation} \label{eq: path loss}
    L(k,GS) = \left(\frac{4\pi f_c d(k,GS)}{c}\right)^2,
\end{equation}
where $f_c$ is the carrier frequency, $c$ is the speed of light, and $d(k,GS)$ is the distance between satellite $k$ \nas{and the GS}.
\nas{M}aximum achievable data rate for \nas{satellite $k$} under the Gaussian channel assumption is 
\begin{equation} \label{eq: rate}
    R(k,GS) = B \log_2 \left(1+\mathrm{SNR}(k,GS)\right).
\end{equation}
We use the longest distance between each satellite and the GS in each on-time duration to derive the SNR and rate.
The time for exchanging the model parameters $\pmb{w}$ between satellite $k$ and the GS then is
\begin{equation} \label{eq: Time for transmitting}
    t_c(k, GS)= \frac{S(\pmb{w})}{R(k,GS)} + \frac{d(k,GS)}{c},
\end{equation}
where \nas{$\frac{S(\pmb{w})}{R(k,GS)}$} and $\frac{d(k,GS)}{c}$ are the required time for transmission and propagation, respectively, and $S(\pmb{w})$ is the data size of $\pmb{w}$ in bits.
\vspace{-0.1 cm}
\section{The Proposed Scheduling Algorithm}

As mentioned above, \nas{a satellite can communicate with the GS when there is a line of sight link between them. We model this as} satellite $k$ is in the $E_\mathrm{on}$ state. As a noteworthy fact, the rotation of Earth causes duration between visits of a satellite to the same GS to be different from its orbital period $T_p$.

Federated Averaging (FedAvg) algorithm is a well-known and widely employed FL procedure \cite{mcmahan2017communication,li2020unified}. Using it to train a FL model on satellites with full client participation~\cite{razmi2021board} \nas{roughly} works as follows: 1) The GS transmits the global
model parameters to \nas{all} satellites when they visit; 2) \nas{S}atellites train the
model using local SGD; 3) \nas{S}atellites send the updated
local parameters to the GS upon their next visit; and 4) The GS aggregates
\nas{received} model parameters from all satellites.

Implementing FedAvg in ground-assisted FL on satellites scenarios leads to very slow model convergence because \nas{satellites} visit the GS at different times and the GS has to wait for all updates to \nas{receive} before starting a new global epoch. An asynchronous version of \nas{FedAvg} algorithm, named FedSat, is proposed in \cite{9674028} for the satellite scenarios and shown to significantly reduce the convergence time. In FedSat, the GS updates the global model parameters whenever it receives updated local parameters from one of the \nas{satellites.}

In this paper, we propose a general approach that helps in implementing the FL for any form of satellite constellation. This approach, as shown in Fig. \ref{fig: Schedule diagram for satellites}, consists of three consecutive steps. The inputs are the satellites and the GS information such as the number of satellites and their altitude, inclination, and initial positions, plus the position of the GS.

\begin{figure}
    \centering
    \includegraphics[width=0.40\textwidth]{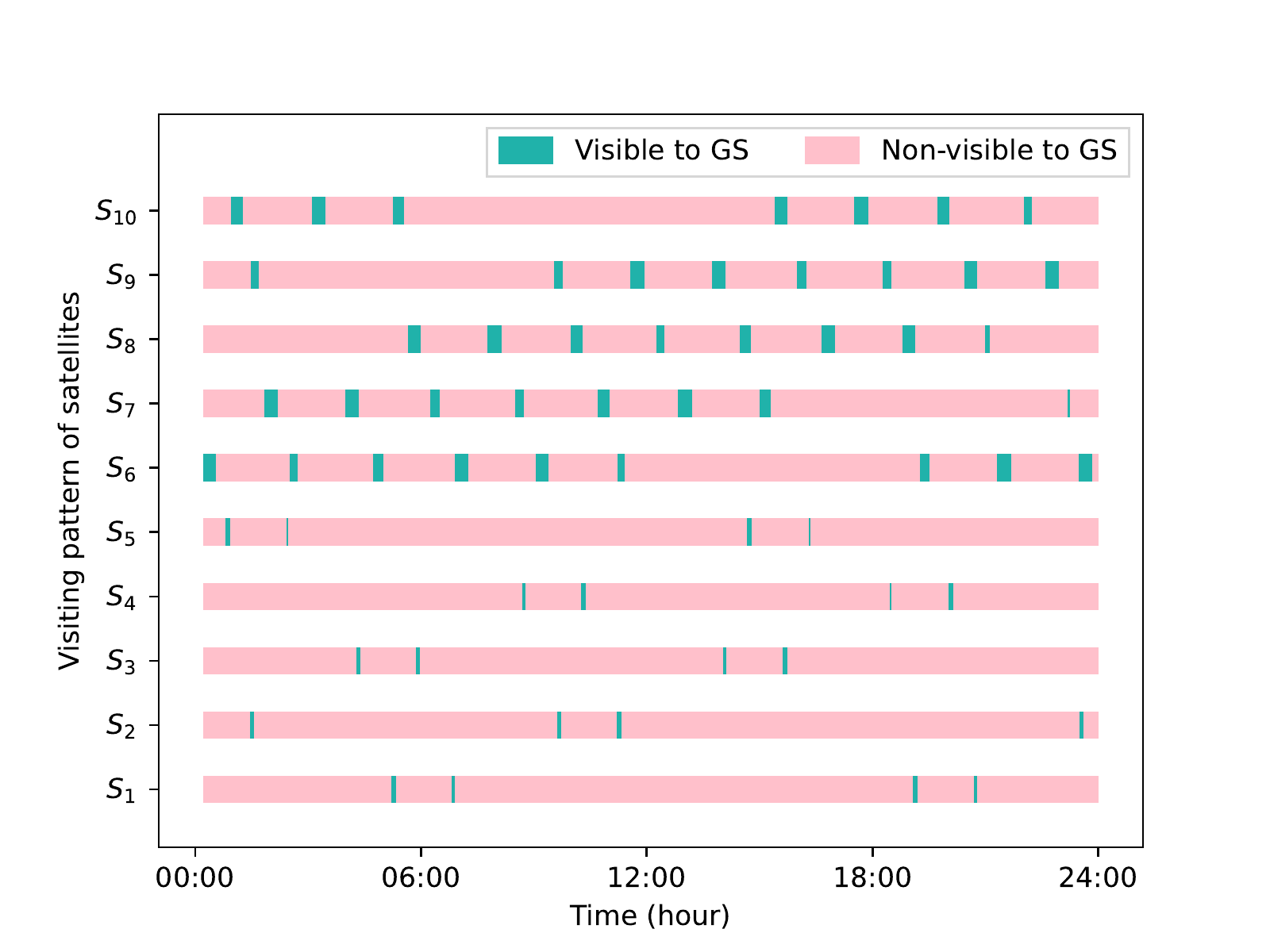}
    \caption{Visiting pattern of 10 satellites and the GS in Bremen in one day. Satellites $S_1$ to $S_5$ are at altitude 500 km and $S_6$ to $S_{10}$ are at altitude 2000 km.}
    \label{fig: visiting pattern}
\vspace{-0.2 cm}    
\end{figure}

With this input data, in the first step, the visiting pattern between each satellite and the GS can be obtained in the considered time i.e. $[T_b,T_f]$. An example of this visiting pattern is illustrated in Fig.~\ref{fig: visiting pattern}. This figure presents the visiting pattern for a period of 24-hour between the GS, located in Bremen, and ten satellites. Five of the satellites, $S_1$ to $S_5$, are at altitude 500~km and the other five, $S_6$ to $S_{10}$, are at altitude 2000~km. The \nas{rise-time}, $\tau_{\mathrm{rise}}$, set-time, $\tau_{\mathrm{set}}$, on-time, $\tau_{\mathrm{on}}$, and off-time, $\tau_{\mathrm{off}}$, of all satellites are derived in this step. Let us define the visiting pattern, $\mathcal{VP}$, as
 \vspace{-0.1 cm}
\begin{equation}
    \mathcal{VP} = \left( \tau_{\mathrm{rise}}, \tau_{\mathrm{set}} \right).
    \vspace{-0.1 cm}
\end{equation}

\begin{figure}[t]
    \centering
    \includegraphics[width=8cm,height=9cm]{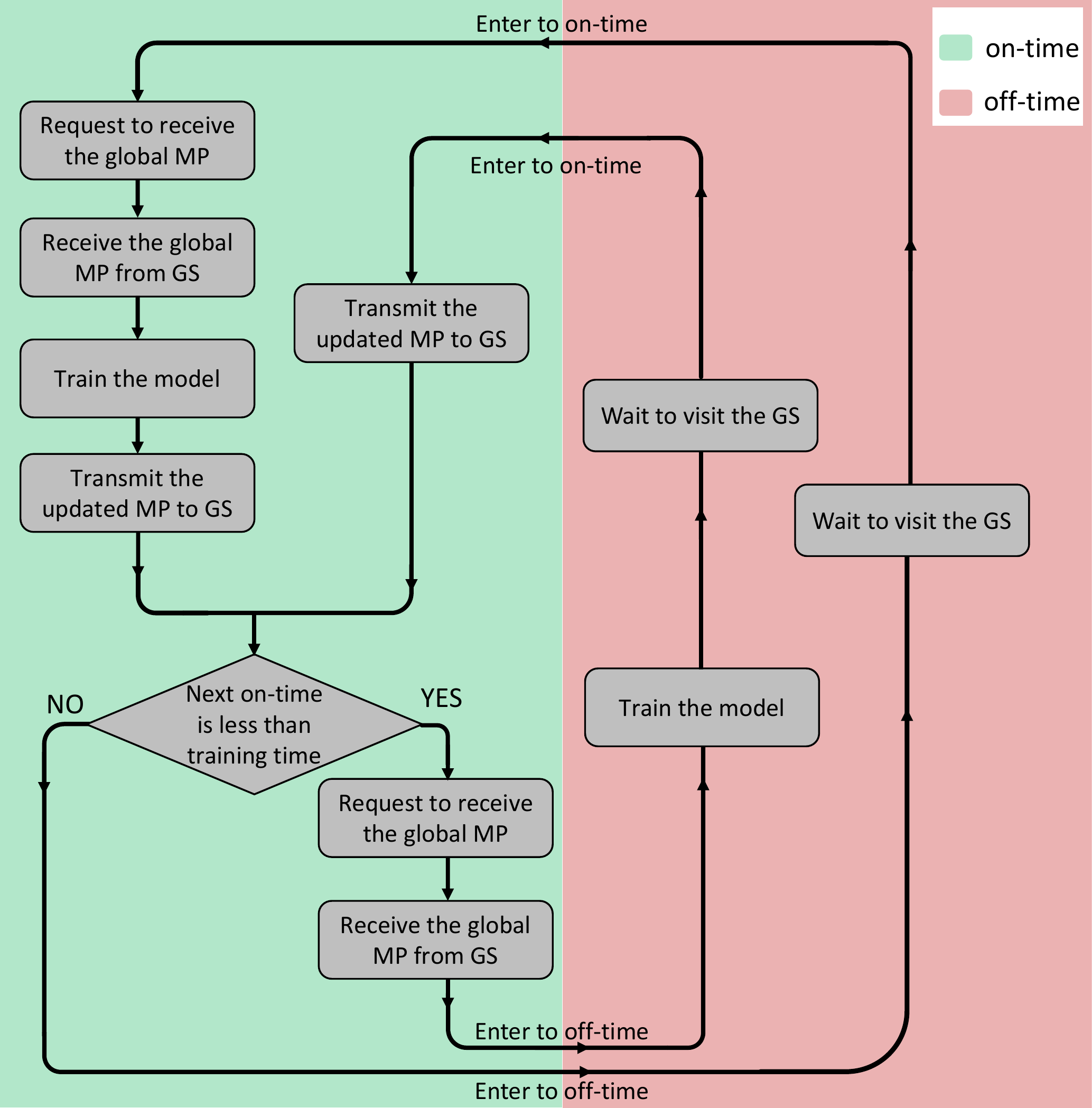}
    \vspace{-0.2 cm}
    \caption{Flow chart of FedSatSchedule algorithm, the red and green colors represent the off-time and on-time intervals, respectively. MP stands for model parameters.}
    \label{fig: Flow chart}
    \vspace{-0.4 cm}
\end{figure}

Then, in the second step, a scheduling algorithm is designed based on the derived $\mathcal{VP}$. For example, the algorithm that will be proposed in Section \ref{section: FedSatSchedule} \nas{uses} $\mathcal{VP}$ to determine whether the satellite trains the next model iteration while being offline or during its next visit to the GS. This is illustrated in Fig.~\ref{fig: Flow chart}.
The scheduling algorithm \nas{in} the second step \nas{leads} to determining the transmission times between the satellites and the GS in the third step, i.e.,
the time intervals in which the UL and DL transmissions to exchange the model parameters happen are extracted. Let us define the sequence of transmission referring to these time intervals as

\vspace{-0.2 cm}
\begin{equation}
    \mathcal{ST} = \left(\tau_{\mathrm{UL}}, \tau_{\mathrm{DL}}\right),\\
\end{equation}
where $\tau_{\mathrm{UL}}$ and $\tau_{\mathrm{DL}}$ are tuples specified by 
\begin{equation}
    \tau_{\mathrm{UL}} = \left(t_{u,1}, t_{u,2}, ..., t_{u,K}\right),\\
\end{equation}
\begin{equation}
    \tau_{\mathrm{DL}} = \left(t_{d,1}, t_{d,2}, ..., t_{d,K}\right),\\
\end{equation}
where $t_{u,k}$ and $t_{d,k}$ are sequences of the UL and DL transmission times associated with the $k$-{th} satellites, given by
\begin{equation} \label{eq: UL sequence}
    t_{u,k} = \left\{t_{u,k}^{n}\right\}_{n=1}^{U_k},
\end{equation}
\begin{equation} \label{eq: DL sequence}
    t_{d,k} = \left\{t_{d,k}^{n}\right\}_{n=1}^{D_k}.\\
\end{equation}
In (\ref{eq: UL sequence}) and (\ref{eq: DL sequence}), $U_k$ and $D_k$ stand for the total number of UL and DL transmissions of the $k$-{th} satellite, respectively.
To obtain the optimal $\mathcal{ST}$, we formulate an optimization problem expressed as
\vspace{-0.2 cm}
\begin{equation} \label{eq:ST}
	\mathcal{ST^*} = \argmax_{\mathcal{ST}} C(\mathcal{VP}, \mathcal{ST}),
\end{equation}
where $C$, as a function of $\mathcal{VP}$ and $\mathcal{ST}$, is a desired design criterion which should be defined based on the requirements of any specific problem. An example of this criterion function is given in section \ref{section: FedSatSchedule}.

By using the proposed three-step thorough model, in the following,
we present a new scheduling algorithm named as FedSatSchedule. To define this scheme, at first, we briefly explain FedSat, the scheme that we proposed in our previous work~\cite{9674028}.
\vspace{-0.2 cm}
\subsection{Federated Learning for Satellite Constellations (FedSat)}

One way to implement \nas{FL} for the satellite \nas{constellations}, is using an asynchronous algorithm as presented in FedSat~\cite{9674028}. \nas{By this approach}, we can benefit from the predictability of satellites visiting pattern \nas{which helps to} overcome the intermittent connectivity between the GS \nas{and satellites.}

In FedSat, each satellite exchanges the model parameters with the GS when they visit each other. This means in the rise-time, satellite $k$ transmits the updated local model parameters to the GS. Then, the GS updates global model parameters by
\begin{equation} \label{eq:fedsat}
	\pmb{w}^{n+1} = \pmb{w}^{n} - \alpha_k (\pmb{w}^{n_k-1,I}_k - \pmb{w}_k^{n_k,I}).
\end{equation}
where $\alpha_k$ is $\frac{D_k}{D}$. Then, the GS transmits the updated model parameters to that satellite. Again, the satellite trains the model in the off-time period and transmits the model parameters to the GS in the next rise-time. This algorithm does not take on-time and off-time duration\nas{s} into account. However, if the satellite's next visit to the GS will be long enough to complete the training during that visit, obtaining the global model already at the current visit will lead to considerable model staleness, which has a negative impact on convergence. Exploiting this simple observation is the key idea behind the FedSatSchedule algorithm proposed next.

\subsection{Federated Learning Scheduling for Satellite Constellations (FedSatSchedule)} \label{section: FedSatSchedule}
In \nas{FedSat} scheme, as mentioned above, the duration of each visit i.e. $t_{\mathrm{on},k}$ is not taken into account when deriving $t_{u,k}$ and $t_{d,k}$ for $\mathcal{ST}$. However, due to the fact that the length of $t_{\mathrm{on},k}$ and $t_{\mathrm{off},k}$ are completely predictable, $\mathcal{ST}$ can be determined such that a higher training accuracy can be achieved in a shorter time frame. The FedSatSchedule scheme \nas{uses} these times to schedule the FL aimed at convergence time reduction. Formalizing this in our general framework, (\ref{eq:ST}) can be converted to
\begin{equation} \label{eq:ST_convergence_problem}
	\mathcal{ST^*} = \argmin_{\mathcal{ST}} CT(\mathcal{VP}, \mathcal{ST})
\end{equation}
where $CT$ is the convergence time of the model which, in its turn, is a function of $\mathcal{ST}$ and $\mathcal{VP}$. Solving this problem exactly is challenging, as even the functional relation $CT$ is difficult to define. Instead, we take a heuristic approach that aims to reduce the model staleness at the satellites while still ensuring that every satellite provides a model update during each visit to the GS. In particular, the scheduler predicts whether the next visit to the GS is long enough to complete a local model update. If this is the case, the satellite will receive the current global model parameters upon \nas{its} next contact to the GS. Otherwise, it will receive them immediately and compute its update during its off-time. We design this procedure named "FedSatSchedule" explicitly next.

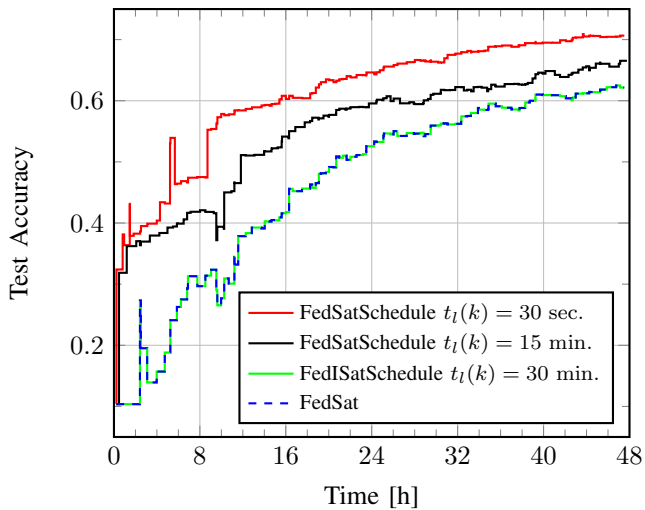
\begin{figure}
\begin{tikzpicture}
\begin{axis}[
thick,
yminorgrids = true, 
legend entries = {FedSatSchedule $t_l(k) = 30~\mathrm{sec}.$, FedSatSchedule $t_l(k) = 15~\mathrm{min}.$, FedISatSchedule $t_l(k) = 30~\mathrm{min}.$, FedSat},
xlabel={Time [h]},
ylabel={Test Accuracy},
grid=major,
grid=minor,
xtick = {0,8,...,48},
xmin = 0,
xmax = 48,
ymin = 0.05,
ymax = 0.75,
minor x tick num = 3,
minor y tick num = 1,
grid = major,
width=1*\axisdefaultwidth,
height=1*\axisdefaultheight,
legend pos=south east,
legend cell align=left,
legend style={font=\footnotesize}
]

\addplot[color=red] table [x=time, y=fedavgPlannedAsync_withoutaug_30_sec, col sep=comma] {Eusipco_Non_IID_Bremen.dat};

\addplot[color=black] table [x=time, y=fedavgPlannedAsync_withoutaug_15_min, col sep=comma] {Eusipco_Non_IID_Bremen.dat};

\addplot[color=green] table [x=time, y=fedavgPlannedAsync_withoutaug_30_min, col sep=comma] {Eusipco_Non_IID_Bremen.dat};

\addplot[color=blue, dashed] table [x=time, y=fedavgPlannedAsync_withoutaug_30_min, col sep=comma] {Eusipco_Non_IID_Bremen.dat};

\end{axis}
\end{tikzpicture}
\vspace{-2.5ex}
\caption{Test Accuracy of a LEO constellation with 10 satellites and a GS located in Bremen.}
\label{fig: FedSatSchedule_results}
\end{figure}

In FedSatSchedule algorithm, in the current on-time i.e. $[t_{r,k}^{n},t_{s,k}^{n}]$, the $k$-th satellite decides about the required operations based on comparing the duration of the next on-time and the necessary time for training; whether \nas{$t_{s,k}^{n+1} - t_{r,k}^{n+1} < t_l(k)$} or not. The flow chart in Fig.~\ref{fig: Flow chart}, in detail shows the tasks to be done during the $n$-{th} on-time period.

If the next on-time period, $t_{s,k}^{n+1} - t_{r,k}^{n+1}$, is shorter than the required training time, $t_l(k)$, the satellite requests that the GS \nas{sends} the global model parameters in the same visit i.e. ($n$)-{th} on-time period. Then, the satellite by using the received global parameters, trains the model in the coming off-time period i.e. $[t_{s,k}^{n},t_{r,k}^{n+1}]$. Afterwards, in the ($n+1$)-{th} on-time interval, it transmits the updated parameters to the GS.

Instead, if the next on-time period is longer than the required time for training, the satellite will have enough time for training using more up-to-date parameters in the coming on-time interval. Note that, in the off-time interval, the GS keep\nas{s} updating the model parameters based on the received parameters from other satellites. Then, the \nas{$k$-th} satellite had better wait and receive up-to-date model parameters exactly before starting to train in the next on-time interval. Hence, the satellite instead of requesting for receiving the new model parameters in the $n$-{th} on-time interval, will do \nas{it} in the ($n+1$)-{th} on-time. With the received parameters, the satellite trains the model and transmits the updated model parameters to the GS in the ($n+1$)-{th} on-time. This approach results in higher accuracy without adding more delay or using extra resources.

\vspace{-0.2 cm}
\section{Numerical Results}
\vspace{-0.1 cm}
In this section, we present simulation results to show the effectiveness of the proposed scheme. We consider ten satellites in 10 orbits; five of them are at altitude 500 km and the other five are at altitude 2000 km with a GS located in Bremen. The minimum difference in right ascension
of the ascending node (RAAN) between two near orbits of different altitudes is 36°. The inclination and minimum elevation angles of all satellites are set to \ang{80} and \ang{10}, respectively.
All satellites and the GS transmit the model parameters on channels with bandwidth of 20 MHz with the transmission power set to 40 $\si{dBm}$. The transmit and receive antenna gains are both set to 6.98 $\si{dBi}$. The carrier frequency and the receiver noise temperature are $f_c = 2.4  \si{GHz}$ and $T = 290~\si{K}$, respectively.

For training process based on \cite{he2020fedml}, the well-known CIFAR dataset with the ResNet-18 model is considered. The learning rate, $\eta$, and the batch sizes are set to 0.1 and 10, respectively. The whole CIFAR dataset is divided between all satellites with Non-IID settings such that five labels are given to the satellites at altitude 500 and the other five labels to the other five at altitude 2000 km.

We examine the impact of the training time of each satellite, $t_l(k)$, on the test-accuracy. 
Fig.~\ref{fig: FedSatSchedule_results} shows the test accuracy for three different training time, 30 seconds, 15 minutes and 30 minutes, for a period of two days. It depicts that our proposed scheduling algorithm can noticeably improve the test accuracy for the cases $t_l(k) = 30$ seconds and $t_l(k) = 15$ minutes compared to the FedSat. 

We observe if $t_l(k) = 30$ seconds, it takes \nas{48} hours for the \nas{FedSat} to have a test accuracy around $62\%$, while for the \nas{FedSatSchedule}, it takes \nas{only 16} hours, \nas{improving the convergcne speed by a factor of three}. FedSatSchedule outperforms FedSat due to a proper scheduling to receive more up-to-date model parameters.

By increasing the training time interval, as we see in the case with the $t_l(k) = 30$ minutes, the \nas{performances} of the FedSat and FedSatSchedule \nas{converge} together. \nas{In} such cases, all satellites have, in practice, to train the model in their off-time period. So, the FedSatSchedule can not benefit from having more up-to-date model parameters.
\vspace{-0.2 cm}
\section{Conclusion}
\vspace{-0.1 cm}
In this paper, we have presented a general approach for optimally scheduling the transmission and reception time of the model parameters between the satellites and the GS for \nas{implementing} FL in any constellation. Then, we have specifically designed a scheduling algorithm, FedSatSchedule, by considering the duration of each on-time. The numerical results have shown that this scheme can accelerate the convergence of FL.
\vspace{-0.3 cm}
\bibliographystyle{IEEEtran}
\bibliography{references}

\begin{thebibliography}{10}
\providecommand{\url}[1]{#1}
\csname url@samestyle\endcsname
\providecommand{\newblock}{\relax}
\providecommand{\bibinfo}[2]{#2}
\providecommand{\BIBentrySTDinterwordspacing}{\spaceskip=0pt\relax}
\providecommand{\BIBentryALTinterwordstretchfactor}{4}
\providecommand{\BIBentryALTinterwordspacing}{\spaceskip=\fontdimen2\font plus
\BIBentryALTinterwordstretchfactor\fontdimen3\font minus
  \fontdimen4\font\relax}
\providecommand{\BIBforeignlanguage}[2]{{%
\expandafter\ifx\csname l@#1\endcsname\relax
\typeout{** WARNING: IEEEtran.bst: No hyphenation pattern has been}%
\typeout{** loaded for the language `#1'. Using the pattern for}%
\typeout{** the default language instead.}%
\else
\language=\csname l@#1\endcsname
\fi
#2}}
\providecommand{\BIBdecl}{\relax}
\BIBdecl

\bibitem{5G-access}
I.~Leyva-Mayorga, B.~Soret, M.~Röper, D.~Wübben, B.~Matthiesen, A.~Dekorsy,
  and P.~Popovski, ``{LEO} small-satellite constellations for {5G} and
  beyond-{5G} communications,'' \emph{IEEE Access}, vol.~8, pp.
  184\,955--184\,964, 2020.

\bibitem{9378798}
J.~M. Haut, M.~E. Paoletti, S.~Moreno-Álvarez, J.~Plaza, J.-A. Rico-Gallego,
  and A.~Plaza, ``Distributed deep learning for remote sensing data
  interpretation,'' \emph{Proceedings of the IEEE}, vol. 109, no.~8, pp.
  1320--1349, 2021.

\bibitem{curzi2020large}
G.~Curzi, D.~Modenini, and P.~Tortora, ``Large constellations of small
  satellites: A survey of near future challenges and missions,''
  \emph{Aerospace}, vol.~7, no.~9, p. 133, 2020.

\bibitem{mcmahan2017communication}
H.~B. McMahan, E.~Moore, D.~Ramage, S.~Hampson, and B.~Aguera~y Arcas,
  ``Communication-efficient learning of deep networks from decentralized
  data,'' ser. Proc. Mach. Learn. Res. (PMLR), vol.~54, 2017.

\bibitem{9674028}
N.~Razmi, B.~Matthiesen, A.~Dekorsy, and P.~Popovski, ``Ground-assisted
  federated learning in leo satellite constellations,'' \emph{IEEE Wireless
  Communications Letters}, pp. 1--1, 2022.

\bibitem{razmi2021board}
{Razmi, Nasrin and Matthiesen, Bho and Dekorsy, Armin and Popovski, Petar},
  ``On-board federated learning for dense {LEO} constellations,'' \emph{arXiv
  preprint arXiv:2111.12769}, 2021.

\bibitem{so2022fedspace}
J.~So, K.~Hsieh, B.~Arzani, S.~Noghabi, S.~Avestimehr, and R.~Chandra,
  ``Fedspace: An efficient federated learning framework at satellites and
  ground stations,'' \emph{arXiv preprint arXiv:2202.01267}, 2022.

\bibitem{8737464}
N.~H. Tran, W.~Bao, A.~Zomaya, M.~N.~H. Nguyen, and C.~S. Hong, ``Federated
  learning over wireless networks: Optimization model design and analysis,'' in
  \emph{IEEE INFOCOM 2019 - IEEE Conference on Computer Communications}, 2019,
  pp. 1387--1395.

\bibitem{ippolito2017satellite}
L.~J. Ippolito~Jr, \emph{Satellite Communications Systems Engineering}.\hskip
  1em plus 0.5em minus 0.4em\relax John Wiley \& Sons, 2017.

\bibitem{li2020unified}
Z.~Li and P.~Richt{\'a}rik, ``A unified analysis of stochastic gradient methods
  for nonconvex federated optimization,'' \emph{arXiv preprint
  arXiv:2006.07013}, 2020.

\bibitem{he2020fedml}
C.~He, S.~Li, J.~So, X.~Zeng, M.~Zhang, H.~Wang, X.~Wang, P.~Vepakomma,
  A.~Singh, H.~Qiu \emph{et~al.}, ``Fedml: A research library and benchmark for
  federated machine learning,'' \emph{arXiv preprint arXiv:2007.13518}, 2020.

\end{thebibliography}
\vspace{-0.3 cm}
\end{document}